# Inferring multiple consensus trees and supertrees using clustering: a review

Vladimir Makarenkov[1*], Gayane S. Barseghyan[1] and Nadia Tahiri[2]

[1] Département d'Informatique, Université du Québec à Montréal, Case postale 8888, Succursale Centre-ville, Montreal, QC, H3C 3P8, Canada
[2] Département d'Informatique, Université de Sherbrooke, 2500 Boulevard de l'Université, Sherbrooke, Québec J1K 2R1, Canada
*Corresponding author: makarenkov.vladimir@uqam.ca

**Abstract:** Phylogenetic trees (i.e. evolutionary trees, additive trees or *X*-trees) play a key role in the processes of modeling and representing species evolution. Genome evolution of a given group of species is usually modeled by a species phylogenetic tree that represents the main patterns of vertical descent. However, the evolution of each gene is unique. It can be represented by its own gene tree which can differ substantially from a general species tree representation. Consensus trees and supertrees have been widely used in evolutionary studies to combine phylogenetic information contained in individual gene trees. Nevertheless, if the available gene trees are quite different from each other, then the resulting consensus tree or supertree can either include many unresolved subtrees corresponding to internal nodes of high degree or can simply be a star tree. This may happen if the available gene trees have been affected by different reticulate evolutionary events, such as horizontal gene transfer, hybridization or genetic recombination. Thus, the problem of inferring multiple alternative consensus trees or supertrees, using clustering, becomes relevant since it allows one to regroup in different clusters gene trees having similar evolutionary patterns (e.g. gene trees representing genes that have undergone the same horizontal gene transfer or recombination events). We critically review recent advances and methods in the field of phylogenetic tree clustering, discuss the methods' mathematical properties, and describe the main advantages and limitations of multiple consensus tree and supertree approaches. In the application section, we show how the multiple supertree clustering approach can be used to cluster aaRS gene trees according to their evolutionary patterns.

# Introduction

The term phylogeny (i.e. phylogenetic tree or evolutionary tree) was introduced by Haeckel in 1866, who defined it as "the history of the paleontological development of organisms by analogy with ontogeny or the history of individual development". A phylogenetic tree represents a hypothesis about evolution of a given group of species which are usually associated with the tree leaves.

In mathematics, phylogenetic trees are called additive trees or *X*-trees (as their leaves are often associated with the set of species *X*; Barthélemy and Guénoche 1991). Let us now present some necessary mathematical definitions related to phylogenetic trees. The distance $\delta(x,y)$ between two vertices *x* and *y* in a phylogenetic tree *T* is defined as the sum of the edge lengths in the unique path linking *x* and *y* in *T*. Such a path is denoted *(x,y)*. A leaf is a vertex of degree one. Usually, a leaf represents a contemporary species (or taxa).

**Definition 1.** *Let X be a finite set of n taxa. A dissimilarity d on X is a non-negative function on X × X such that for any x, y from X:*
$d(x,y) = d(y,x) \geq d(x,x) = 0$.

**Definition 2.** *A dissimilarity d on X satisfies the four-point condition if for any x, y, z, and w from X: $d(x,y) + d(z,w) \leq \max\{d(x,z) + d(y,w); d(x,w) + d(y,z)\}$.*

**Definition 3.** *For a finite set X, a phylogenetic tree (i.e. an additive tree or an X-tree, i.e. a tree whose leaves are labeled according to a final set of species X) is an ordered pair $(T, \varphi)$ consisting of a tree T, with vertex set V, and a map $\varphi: X \to V$ with the property that, for all $x \in X$ with degree at most two, $x \in \varphi(X)$. A phylogenetic tree is binary if $\varphi$ is a bijection from X into the leaf set of T and every interior vertex has degree three.*

The theorem relating the four-point condition and a dissimilarity representability by a phylogenetic tree is as follows:

**Theorem 1** *(Zarestskii, Buneman, Patrinos & Hakimi, Dobson). Any dissimilarity satisfying the four-point condition on X × X (where X is a finite set of species) can be represented by a phylogenetic tree T such that for any x, y from X, d(x,y) is equal to the length of the path linking the leaves x and y in T. This dissimilarity is called a tree metric. Furthermore, this tree is unique.*

Figure 1 gives an example of a tree metric on the set *X* of five taxa and the corresponding phylogenetic tree.

Unfortunately, real-life evolutionary distances (or dissimilarities) rarely satisfy the four-point condition. Thus, one need to carry out an approximation algorithm to infer a tree metric matrix from a given matrix of evolutionary distances (Gascuel 2005). Among the most known distance-based approximation algorithms we can mention Neighbor-Joining (Saitou 1988), UPGMA (Sokal and Michener 1958), FITCH (Felsenstein 1997), and MW (Makarenkov and Leclerc 1996, 1999).

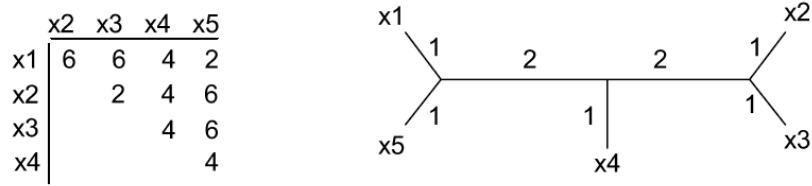

**Fig 1.** *An example of a tree metric on the set X of five taxa (on the left) and the corresponding phylogenetic tree (additive tree or X-tree) on the right.*

Biologists often need to compare phylogenetic trees to each other in order to discover different evolutionary histories that govern a given set of species. There are several measures for comparing phylogenetic trees. The most popular of them include the Robinson and Foulds topological distance (*RF*) (Robinson and Foulds 1981), the least-squares distance (*LS*), the bipartition dissimilarity (*BD*) (Boc et al. 2010), and the quartet distance (*QD*) (Bryant et al. 2000). In this literature review, we will mainly explore the methods based on the Robinson and Foulds distance. The Robinson and Foulds topological distance (Robinson and Foulds 1981) between two trees is the minimum number of elementary operations (contraction and expansion) of nodes needed to transform one phylogenetic tree into another. It is also the number of splits (or bipartitions) that are present in one tree and absent in the other. The two phylogenetic trees in question must have the same set of taxa. The closer two phylogenetic trees are topologically, the smaller the value of the *RF* distance. It is often relevant to normalize the value of the *RF* distance by dividing it by its maximum possible value (equal to 2*n*-6) for two binary phylogenetic trees with *n* leaves. The *RF* distance calculation between two trees with *n* leaves can be carried out in *O*(*n*) (Day 1985, Makarenkov 1997, Makarenkov and Leclerc 2000).

Often phylogenetic tree reconstruction methods do not return a single phylogenetic tree as output, but a collection of different trees (Gascuel 2005). Moreover, phylogenetic trees inferred for different genes often differ from each other. There is no absolute criterion for determining whether one tree is better than the others (except for the use of intrinsic criteria, e.g., the use of bootstrap scores). For this reason, it is preferable to seek a consensus representation of these trees, such that their concordant parts appear clearly in relation to the discordant parts. The resulting representation is called a *consensus tree*. Traditional consensus methods generate a single phylogenetic tree that is a representative of all of the input trees (Bryant 2003). One of the first consensus methods was proposed by Adams (Adams 1972). Since then, a wide variety of methods have been developed. How to use them has been the subject of much debate (Bryant 2003, Dong et al. 2010).

The main types of consensus trees are the following: the strict consensus tree (Sokal and Rohlf 1981, Moon and Eulenstein 2017), the majority-rule consensus tree (Margush and McMorris 1981), the Nelson consensus tree (Nelson 1979), and the extended majority-rule consensus tree (Felsenstein 1985). Let us briefly recall the main characteristics of each of these consensus trees.

The *strict consensus tree* (or Nelson's cladogram) is inferred by considering only those tree splits (i.e. bipartitions induced by the internal tree edges) that are identical in all trees compared. Conflicting parts of phylogenetic trees are represented by multifurcations in a strict consensus tree.

It is sometimes more convenient to have a less strict criterion than the one used by the strict consensus tree in order to allow bipartitions that are not necessarily present in all trees. When comparing a set of phylogenetic trees with different topologies, it is possible to search for the monophyletic groups that appear most frequently (often in more than 50% of the trees) among all the trees compared. The resulting tree is the *majority-rule consensus tree*.

The *extended majority-rule consensus tree* contains all majority bipartitions to which the remaining compatible bipartitions are added in turn, starting with the most frequent bipartitions for the given tree set. The process stops when a completely resolved (i.e. binary) tree is obtained. The extended majority consensus tree is the most frequently used in molecular biology, as it is always the best resolved among the three types of consensus trees discussed so far.

The *Nelson consensus tree* includes the heaviest set of compatible bipartitions. It consists in finding a clique of maximum weight in a compatibility graph of the entire bipartition set, which is NP-hard (Nelson 1979, Bryant 2003).

Unfortunately, in many practical situations, phylogenetic trees used as input of consensus tree reconstruction methods can be quite divergent. This can happen, for example, when the input trees represent the evolution of different genes which have been affected by multiple reticulate evolutionary events such as horizontal gene transfer, hybridization or intragenic/intergenic recombination, ancient gene duplication or gene loss (Makarenkov and Legendre 2000, Mirkin et al. 2003, Bapteste et al. 2004). These evolutionary events can be unique for a subgroup of the input gene trees. Thus, it seems to be much more appropriate to represent this subgroup by its own consensus tree. However, the conventional consensus tree methods provide only one candidate tree for a given set of input gene phylogenies without considering their possible subgroups (or clusters) (Maddison et al. 2007).

Figure 2 shows an example of four seven-leaf phylogenetic trees $T_1$, $T_2$, $T_3$, and $T_4$. Here, the solution consisting of two majority-rule consensus trees, $T_{12}$ and $T_{34}$, seems to be much more appropriate than the conventional consensus solution consisting of a single majority-rule consensus tree, $T_{1234}$, i.e., here a star tree (a tree having no internal edges at all).

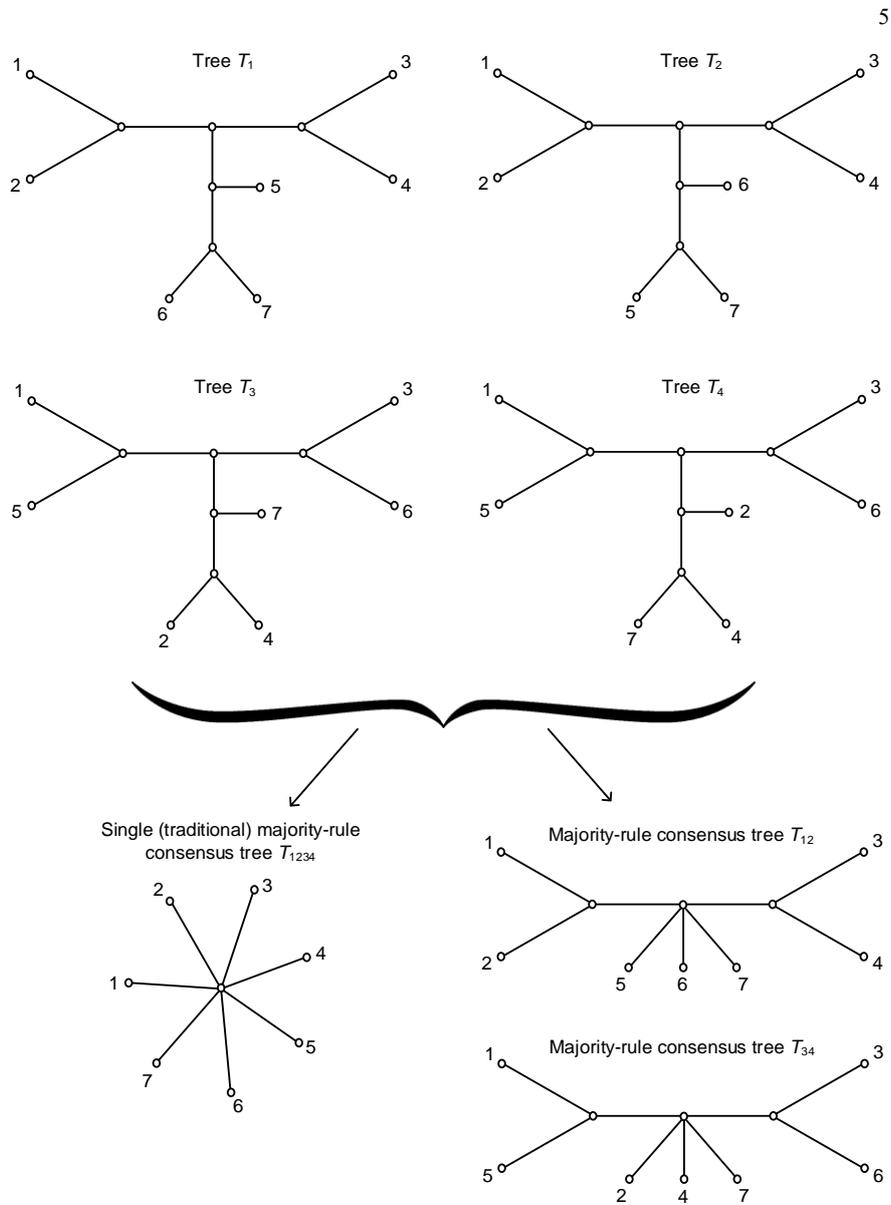

**Fig 2.** *Four phylogenetic trees $T_1$, $T_2$, $T_3$, and $T_4$ defined on the same set of seven leaves. Their single (traditional) majority-rule consensus tree is a star tree $T_{1234}$. The majority-rule consensus trees, $T_{12}$ and $T_{34}$, constructed for the pairs of topologically close trees: $T_1$ and $T_2$, and $T_3$ and $T_4$, respectively.*

In many evolutionary studies gene trees to be combined are defined on different, but partially overlapping, sets of taxa (e.g. see Tree of Life project; Maddison et al. 2007). It is very unlikely that all the genes considered have been sequenced for the same sets of species. In order to reconcile such trees, *supertree reconstruction* methods should be applied (Bininda-Emonds 2004, Wilkinson et al. 2007, McMorris and Wilkinson 2011, Warnow 2018). Supertrees synthesize a given set of small (i.e. partial) trees with partial taxon overlap into comprehensive supertrees that include all taxa present in the given set of trees.

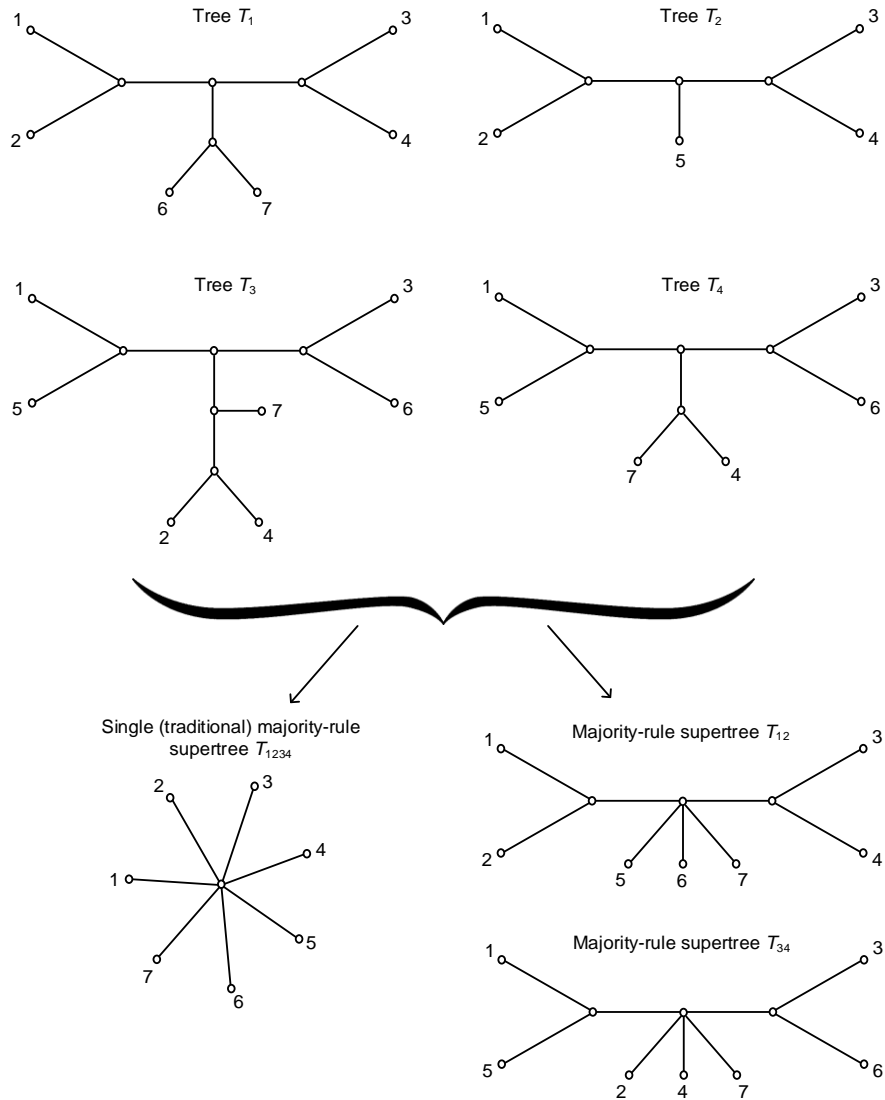

**Fig 3**. *Four phylogenetic trees $T_1$, $T_2$, $T_3$, and $T_4$ defined on different, but mutually overlapping, sets of seven taxa. Their single (traditional) majority-rule supertree is a star tree $T_{1234}$. The majority-rule supertrees, $T_{12}$ and $T_{34}$, constructed for the pairs of topologically close trees: $T_1$ and $T_2$, and $T_3$ and $T_4$, respectively.*

The most known supertree inference method is Matrix Representation with Parsimony (MRP) (Baum 1992, Ragan 1992) that carries out matrix-like aggregation of the given partial trees. The supertree reconstruction methods are commonly used for phylogenetic analysis of organisms with large genomes (Mank et al. 2005, Bininda-Emonds et al. 2007, Faurby et al. 2016, Kimbal et al. 2019). For organisms with small genomes, such as prokaryotes, several approaches to genomic phylogenetic analysis have been adopted. In particular, supertree analysis provides new insights into the evolution of prokaryotes that could not be solved by many other approaches (Daubin et al. 2001). Recently, Makarenkov et al. (2021) and Tahiri et al. (2022) have used supertree phylogenetic analysis to characterize the evolution of SARS-CoV-2 genes.

As in the case of consensus trees, in many practical situations multiple conservative supertrees should be inferred to best represent the evolution of a given group of gene trees. Figure 3 shows an example of four phylogenetic trees $T_1$, $T_2$, $T_3$, and $T_4$ defined on different, but mutually overlapping, sets of seven taxa. Here, the solution consisting of two majority-rule supertrees, $T_{12}$ and $T_{34}$, is more appropriate than that consisting of a single majority-rule supertree, $T_{1234}$, i.e., here a star tree, yielded by the traditional supertree approach.

The idea of building multiple consensus trees was originally formulated by Maddison (Maddison 1991). He discovered that consensus trees for some subsets of input trees may differ a lot and that they are generally much better resolved than the single traditional consensus tree characterizing the whole set of the input trees. Many approaches have been developed to provide solutions for classifying phylogenetic trees based on the well-known clustering algorithms, such as *k*-means and *k*-medoids. We discuss their main features in the Methods section.

Partitioning is a clustering approach used to divide a given set of elements (or taxa) into a meaningful set of groups of elements (objects or entities) called clusters (or classes) (Mirkin 1996, Mirkin 2005). The objective of partitioning is to find groups of similar elements according to a given similarity measure. The four main partitioning approaches that can be used to group the elements based on the set of their features (or variables) are the following: 1) a center of gravity, i.e., the *k*-means algorithm (Lloyd 1957, MacQueen 1967), where *k* denotes the number of clusters; 2) a geometric median, i.e., *k*-medians (Bradley et al. 1997); 3) a center containing the most frequent modes, i.e., *k*-modes (Huang 1998); 4) a medoid-based approach, in which a medoid is a cluster element that minimizes the sum of the distances between it and all other cluster elements, i.e., *k*-medoids (Kaufman and Rousseeuw 1990). In our literature review, we will mainly focus only on the *k*-means and *k*-medoids algorithms as they have been extensively used in tree clustering (see the Methods section). Both of them are very fast, as the time complexity of *k*-means is $O(I \times K \times M \times N)$, where $I$ is the number of iterations in the internal loop of *k*-means, $K$ is the number of clusters, $M$ is the number of features characterizing the given set of elements, and $N$ is the number of elements, whereas the time complexity of *k*-medoids is $O(I \times K \times M \times (N-K)^2)$. It is worth noting that the *k*-medoids algorithm is much less sensitive to outliers than *k*-means. The Euclidean, Manhattan and Minkowski metrics are the most frequently used in the objective function of *k*- means and *k*-medoids (Mirkin 2005, de Amorim and Mirkin 2012, de Amorim and Makarenkov 2016). However, in the case of tree clustering the Robison and Foulds topological distance or another tree distance should be used instead, and phylogenetic trees will play the role of cluster elements.

## Methods

**The Phylogenetic Islands** (Maddison 1991) is a method that divides a collection of trees based on the branch length of the trees and the number of branch rearrangements by which the input trees differ. The author considers the three following types of branch rearrangement: NNI (nearest neighbor interchange), SPR (subtree pruning-regrafting), and TBR (tree bisection reconnection) (Swofford and Olsen 1990, Swofford 1991). In NNI rearrangements, a clade (i.e. a subtree) can be moved to a nearby branch only, in SPR, it can be moved to a nearby or a distant branch, and in TBR, it can be moved to a nearby or a distant branch, with the clade also being rerooted. This method was developed to find the most-parsimonious trees using tree search algorithms, i.e., it starts with multiple starting points to find multiple islands. Maddison formally defines an *island of trees* of length *L* as a collection of *n* trees that satisfy three requirements: (1) all trees are of length <*L*; (2) each tree is connected to every other tree in the island through a series of trees, all of the length <*L*, with adjacent trees in the series differing only by a single rearrangement; and (3) all trees that satisfy criteria 1 and 2 are included in the set. Multiple islands can be discovered by performing many searches with a tree search tools available in PAUP* (Swofford 1991) and Henning86 (Farris 1988), each search starting with a different tree. The trees are generally much more similar within islands than between islands, as shown by the analysis of partition metrics between trees (e.g. the Robinson and Foulds distance or the partition metric). The author concluded that trees on different islands may have different effects on trait evolution.

**Characteristic trees that minimize the information loss** (Stockham et al. 2002) is an alternative approach to single consensus postprocessing methods in phylogenetic analysis. The presented approach was developed using popular clustering algorithms, namely *k*-means and agglomerative clustering (Maddison 1991). The method proposed by Stockham et al. minimises the information loss using the characteristic tree concept. This method can be used to improve the resolution level of the output consensus trees and to provide more details about how the candidate trees are distributed. The major limitation of this method is that the input phylogenetic trees must have the same set of species (consensus case) and the method cannot address the case of homogeneous data (i.e. when the number of cluster *K* = 1). The objective function of the method considered by Stockham et al. (2002) is as follow:

$$OF = \sum_{k=1}^{K} \sum_{i=1}^{N_k} RF^2(T_k^{st}, T_{ki}), \tag{1}$$

where $K$ is the number of clusters, $N_k$ is the number of trees in cluster $k$, $RF^2$ is the squared Robinson and Foulds topological distance between the tree $T_{ki}$ (i.e. tree $i$ of cluster $k$) and the tree $T_k^{st}$ that is the strict consensus tree of cluster $k$.

**Multipolar Consensus (MPC) method** (Bonnard et al. 2006) consists of finding a small set of trees including all splits with support greater than a predefined threshold. Given the splits to be displayed, the number of trees in the multipolar consensus must be minimised. This method can display more secondary evolutionary signals than majority-rule consensus. As the methods of Maddison (1991) and Stockham et al. (2002), the MPC method always generates as a solution multiple consensus trees and never a single one. Bonnard et al. (2006) rely on a heuristic coloring scheme, called Greedy Coloring Algorithm, that uses two main steps: (1) to create an order on the vertices; and (2) to consider the vertices one by one in that order, assigning to a vertex the first color that is not assigned to an already colored vertex related to it. The MPC method differs from the other tree clustering methods in at least two ways: 1) it is more parsimonious, as each non-kernel split present in an input tree is represented only once; 2) it does not require prior clustering of the input trees. As a result, the time complexity of MPC is polynomial on the number of input splits, but only linear on the number of input trees.

**The TreeOfTrees method** (Darlu and Guénoche 2011) allows the comparison of $X$-tree topologies obtained from multiple sets of aligned gene sequences. The main goal of this method is to detect genes with identical histories using bootstrap sampling, and weighted or unweighted consensus. The comparison between tree sets is based on several tree metrics leading to a unique tree labelled by the gene trees (i.e. a kind of hierarchical tree clustering is presented). To estimate the robustness of the congruence between the input gene trees, a resampling procedure is used, which results in the construction of a "tree of gene trees" that provides both: a simple tree representation of the proximity of the gene trees, and a bootstrap value for each bipartition of the *tree of trees*. Each leaf of the tree of trees corresponds to a single gene (or a bootstrapped phylogenetic tree representing its evolution). The comparison between tree topologies starts by transforming each of the input trees into a pairwise distance matrix, counting the number of edges separating two taxa, or using a path length metric. The resulting tree distance matrix allows an unambiguous determination of the tree topology. The consensus tree $T$ is constructed by enumerating all bipartitions belonging to the set of the input trees. Darlu and Guénoche propose the weighted consensus method, defined using the following weight function:

$$w(B_i) = \sum_{T_k \in S} \tau_k, \tag{2}$$

where $B_i$ is the bipartition $i$, $T_k$ is a tree of cluster $k$, $S$ is the subset of the input trees containing the bipartition $B_i$, and $\tau_k$ is the measure of the quality of the tree $T_k$. Then, the authors define the weight of each of the input tree $T_k$ as the sum of the weights of the internal edges contained in $T_k$ using the following formula:

$$\Omega(T_k) = \sum_{B_i \in T_k} w(B_i). \tag{3}$$

**Multiple Consensus Trees** (Guénoche 2013) is a tree clustering method intended to decide whether there is a single consensus among the input gene trees or not, and to detect divergent genes using a partitioning method. If the given gene trees are all congruent, they should be compatible with a single consensus tree. Otherwise, multiple consensus trees corresponding to divergent genetic patterns can be identified. The multiple consensus tree method optimises a generalised score, over a set of tree partitions to decide whether the given set of gene trees is homogeneous or not. The author considers unrooted *X*-trees only and focuses on the following consensus strategies: an *X*-tree is represented by a set of its bipartitions, each corresponding to an internal edge of the tree. Removing each internal edge results in a split, and hence a bipartition of the set of taxa *X*. The weight of each bipartition $B_i = X_i \cup X_i'$ is the number $N_i$ of *X*-trees in the profile that contain that bipartition. The author defines the weight of an *X*-tree *T*, relative to the tree profile $\pi$ of *N* trees, as follows:

$$W_\pi(T) = \sum_{B_i \in T_m} w(B_i) = \sum_{B_i \in T_m} N_i, \tag{4}$$

where $w(B_i)$ is the weight of each bipartition $B_i$ of *T*, $N_i$ is the number of internal majority edges (i.e. the edges satisfying the following condition $N_i > \frac{N}{2}$), and $T_m$ is the tree *T* restricted to its majority edges. The weight of each bipartition $B_i$ is the number $N_i$ of *X*-trees in the profile containing this bipartition.

The author generalizes the score (4), defining it for a partition of trees $P_\pi$ in *k* classes, as follows:

$$W^k(P_\pi) = \sum_{i=1,\ldots,k} p_i \times W_{\pi_i}(T_i^{maj}), \tag{5}$$

where $P_\pi$ is a partition of the set of trees $\pi$ in *k* classes $(\pi_1, \ldots, \pi_k)$ containing respectively $\{p_1, \ldots, p_k\}$ trees, and $T_i^{maj}$ is the majority consensus trees corresponding to class *i*.

**Islands of Trees** (Silva and Wilkinson 2021) is the method based on any appropriate pairwise tree-to-tree distance metric that extends the notion of island to any set or multiset of trees, such as those that can be generated by Bayesian or bootstrap methods and facilitates finding islands of trees *a posteriori*. This can be useful when the strict consensus of most parsimonious trees is relatively unresolved, although it relies on the analytical program (Silva and Wilkinson used PAUP*) to identify not only the number of islands, but also the constituents of most parsimonious trees. Distinct subsets of trees, such as tree islands, are complementary to other means of data exploration that involve attempts at partitioning sets of trees to obtain better summaries and promote better understanding of evolution. However, this method is of limited use for large phylogenetic tree distributions because it replaces the calculation of the distance with a very large number of pairwise comparisons of trees.

**Inferring multiple consensus trees using *k*-medoids** (Tahiri et al. 2018) is a fast method for inferring multiple consensus trees from a given set of phylogenetic trees defined on the same set of species. This method is based on the *k*-medoids partitioning algorithm to partition a given set of trees into multiple tree clusters. The well-known Silhouette and Caliński-Harabasz cluster validity indices have been adapted for tree clustering with *k*-medoids to determine the most appropriate number of clusters. It can be used to identify groups of gene trees that have similar evolutionary histories within the group and different evolutionary histories between the groups. This method is suitable for the analysis of large genomic and phylogenetic datasets.

Compared to the objective function used by Stockham et al. (2002) (see Equation 1), Tahiri et al. (2018) used the majority-rule consensus tree instead of the strict consensus tree, and the unsquared *RF* distances instead of the squared one. The straightforward objective function to be minimized is then as follows:

$$OF = \sum_{k=1}^{K} \sum_{i=1}^{N_k} RF(T_k^{maj}, T_{ki}), \quad (6)$$

where *RF* is the Robinson and Foulds distance between the tree $T_{ki}$ (i.e. tree *i* of cluster *k*) and $T_k^{maj}$ that is the majority-rule consensus tree of cluster *k*. Nevertheless, computing the majority-rule consensus tree or the extended majority-rule consensus tree requires at least $O(nN)$ time, where *n* is the number of leaves (taxa or species) in each tree and *N* is the number of trees.

Thus, Tahiri et al. (2018) used the following objective function in their method which is based on *k*-medoids:

$$OF_{med} = \sum_{k=1}^{K} \sum_{i=1}^{N_k} RF(T_k^m, T_{ki}), \quad (7)$$

where $T_k^m$ is the medoid of cluster *k*, defined as a tree belonging to cluster *k* that minimizes the sum of the *RF* distances between it and all other trees in *k*. This version of the objective function is much faster than that based on Equation (6) because it does not require the majority-rule consensus tree recomputation at each basic step of clustering algorithm. The running time of this method is $O(nN^2+rK(N-K)^2I)$, where $O(nN^2)$ is the time needed to precalculate the matrix of pairwise *RF* distances of size ($N \times N$) between all input trees, *K* is the number of clusters, *I* is the number of iterations in the internal loop of *k*-medoids, and *r* is the number of different random starts used in *k*-medoids (usually hundreds of different random starts are needed to obtain good clustering results; Mirkin 2005).

**Inferring multiple consensus trees and supertrees using *k*-means** (Tahiri et al. 2022) is a new method for inferring multiple alternative consensus trees and supertrees that best represent the main evolutionary patterns of a given set of gene trees. This method is based on the use of the popular *k*-means clustering algorithm and the Robinson and Foulds topological distance. It partitions a given set of trees into one, for homogeneous data, or multiple, for heterogeneous data, cluster(s) of trees. The authors show how the popular Caliński-Harabasz, Silhouette, Ball and Hall, and Gap cluster validity indices can be used in tree clustering with *k*-means. The Euclidean property of the square root of the Robinson and Foulds distance is used to define a fast and efficient objective function that is as follows:

$$OF_{EA} = \sum_{k=1}^{K} \frac{1}{N_k} \sum_{i=1}^{N_k-1} \sum_{j=i+1}^{N_k} RF(T_{ki}, T_{kj}), \qquad (8)$$

The time complexity of the tree clustering algorithm based on Equation (8) is $O(nN^2+rNKI)$.

Moreover, the authors establish some interesting properties, and use them in the clustering process, of the general objective function defined in Equation (6). Specifically, the lower and the upper bounds of this objective function *OF* are established in Theorem 2 below:

**Theorem 2** *(Tahiri et al. 2022). For a given cluster k containing $N_k$ phylogenetic trees (i.e. additive trees or X-trees) the following inequalities hold:*

$$\frac{1}{N_k - 1} \sum_{i=1}^{N_k-1} \sum_{j=i+1}^{N_k} RF(T_{ki}, T_{kj}) \leq \sum_{i=1}^{N_k} RF(T_k^{maj}, T_{ki}) \leq \frac{2}{N_k} \sum_{i=1}^{N_k-1} \sum_{j=i+1}^{N_k} RF(T_{ki}, T_{kj}), \quad (9)$$

*where $N_k$ is the number of trees in cluster k, $T_{ki}$ and $T_{kj}$ are, respectively, trees i and j in cluster k, and $T_k^{maj}$ is the majority-rule consensus tree of cluster k.*

In the same paper, Tahiri et al. show how their method can be extended to the case of supertree clustering. In the supertree clustering context, we assume that a given set of *N* unrooted phylogenetic trees may contain different, but mutually overlapping, sets of leaves. In this case, the original objective function *OF* shown in Equation (6) can be reformulated as follows:

$$OF_{ST} = \sum_{k=1}^{K} \sum_{i=1}^{N_k} RF_{norm}(ST_k, T_{ki}) = \sum_{k=1}^{K} \sum_{i=1}^{N_k} \left( \frac{RF(ST_k, T_{ki})}{2n(ST_k, T_{ki}) - 6} \right), \qquad (10)$$

where *K* is the number of clusters, $N_k$ is the number of trees in cluster *k*, $RF_{norm}(ST_k,T_{ki})$ is the normalized Robinson and Foulds topological distance between tree *i* of cluster *k*, denoted $T_{ki}$, and the majority-rule supertree of this cluster, denoted $ST_k$, reduced to a subtree having all leaves in common with $T_{ki}$. The *RF* distance is normalized here by dividing it by its maximum possible value (i.e. $2n(ST_k,T_{ki})$-6, where $n(ST_k,T_{ki})$ is the number of common leaves in $ST_k$ and $T_{ki}$). The *RF* distance normalization is performed here to account equally the contribution of each tree to clustering. Clearly, Equation (10) can be used only if the number of common leaves in $ST_k$ and $T_{ki}$ is larger than 3.

An analog of Equation (8) can be used in supertree clustering to avoid supertree recalculations at each step of *k*-means. This can be done using the following objective function:

$$OF_{ST\_EA} = \sum_{k=1}^{K} \frac{1}{N_k} \sum_{i=1}^{N_k-1} \sum_{j=i+1}^{N_k} \left( \frac{RF(T_{ki}, T_{kj})}{2n(T_{ki}, T_{kj}) - 6} + \alpha \times \frac{n(T_{ki}) + n(T_{kj}) - 2n(T_{ki}, T_{kj})}{n(T_{ki}) + n(T_{kj})} \right), \quad (11)$$

where $n(T_{ki})$ is the number of leaves in tree $T_{ki}$, $n(T_{kj})$ is the number of leaves in tree $T_{kj}$, $n(T_{ki},T_{kj})$ is the number of common leaves in trees $T_{ki}$ and $T_{kj}$, and $\alpha$ is the penalization (tuning) parameter, taking values between 0 and 1, needed to prevent from putting to the same cluster trees having small percentages of common leaves.

The simulations conducted by Tahiri at al. (2022) illustrated that their new tree clustering method is faster and generally more efficient than the methods of Stockham et al. (2002), Tahiri et al. (2018) and Bonnard et al. (2006) discussed earlier in this section.

**Cluster validity indices adapted to tree clustering**

In this section, we show how the popular Caliński-Harabasz, Silhouette, Ball and Hall, and Gap cluster validity indices can be used in tree clustering with *k*-means.

*Caliński-Harabasz cluster validity index adapted for tree clustering*

The first cluster validity index we consider here is the Caliński-Harabasz index (Caliński and Harabasz 1974). This index, sometimes called the variance ratio criterion, is defined as follows:

$$CH = \frac{SS_B}{SS_W} \times \frac{N-K}{K-1}, \quad (12)$$

where $SS_B$ is the index of intergroup evaluation, $SS_W$ is the index of intragroup evaluation, $K$ is the number of clusters and $N$ is the number of elements (i.e. trees in our case). The optimal number of clusters corresponds to the largest value of *CH*.

In the traditional version of *CH*, when the Euclidean distance is considered, the $SS_B$ coefficient is evaluated by using the $L_2$-norm:

$$SS_B = \sum_{k=1}^{K} N_k \|m_k - m\|^2, \quad (13)$$

where $m_k$ ($k = 1 \ldots K$) is the centroid of cluster $k$, $m$ is the overall mean (i.e. centroid) of all elements in the given dataset $X$, and $N_k$ is the number of elements in cluster $k$. In the context of the Euclidean distance, the $SS_W$ index can be calculated using the two following equivalent expressions:

$$SS_W = \sum_{k=1}^{K} \sum_{i=1}^{N_k} \|x_{ki} - m_k\|^2 = \sum_{k=1}^{K} \frac{1}{N_k} \left( \sum_{i=1}^{N_k-1} \sum_{j=i+1}^{N_k} \|x_{ki} - x_{kj}\|^2 \right), \quad (14)$$

where $x_{ki}$ and $x_{kj}$ are elements *i* and *j* of cluster *k*, respectively (Caliński and Harabasz 1974).

To use the analogues of Equations (13) and (14) in tree clustering, Tahiri et al. (2022) used the concept of centroid for a given set of trees. The median tree (Barthélemy and Monjardet 1981; Barthélemy and McMorris 1986) plays the role of this centroid in a tree clustering algorithm. The median procedure (Barthélemy and Monjardet 1981) is defined below. The set of median trees, Md($\Pi$), for a given set of trees $\Pi = \{T_1, \ldots, T_N\}$ having the same set of leaves $S$, is the set of all trees $T$ defined on $S$, such that: $\sum_{i=1}^{N} RF(T, T_i)$ is minimized. If $N$ is odd, then the majority-rule consensus tree, Maj($\Pi$) of $\Pi$, is the only element of Md($\Pi$). If $N$ is even, then Md($\Pi$) is composed of Maj($\Pi$) and of some more resolved trees.

Tahiri et al. (2022) proposed to use some formulas based on the properties of the Euclidean distance to define $SS_B$ and $SS_W$ in $k$-means-like tree clustering. These formulas do not require the computation of the majority (or the extended majority)-rule consensus trees at each iteration of $k$-means. Precisely, they replace the term $\|x_{ki} - x_{kj}\|^2$ in Equation (14) by $RF(T_{ki}, T_{kj})$ to obtain the formula for $SS_W$:

$$SS_W = \sum_{k=1}^{K} \frac{1}{N_k} \sum_{i=1}^{N_k-1} \sum_{j=i+1}^{N_k} RF(T_{ki}, T_{kj}), \qquad (15)$$

where $T_{ki}$ and $T_{kj}$ are trees $i$ and $j$ of cluster $k$, respectively.

Also, in the case of the Euclidean distance, the formula is as follows:

$$SS_B + SS_W = \frac{1}{N}\left(\sum_{i=1}^{N-1} \sum_{j=i+1}^{N} \|x_i - x_j\|^2\right), \qquad (16)$$

where $x_i$ and $x_j$ are two different elements of $X$ (Caliński and Harabasz 1974).

As a result, the approximation to the global variance between groups, $SS_B$, can be evaluated as follows:

$$SS_B = \frac{1}{N}\left(\sum_{i=1}^{N-1} \sum_{j=i+1}^{N} RF(T_i, T_j)\right) - SS_W, \qquad (17)$$

where $T_i$ and $T_j$ are trees $i$ and $j$ in the set of trees $\Pi$, and $SS_W$ is calculated according to Equation (15).

Based on the Euclidean properties of the square root of the Robinson and Foulds distance, Equations (15) and (17) establish the exact formulas for calculating the indices $SS_B$ and $SS_W$ for the objective function $OF_{EA}$ defined by Equation (8). Interestingly the objective function $OF_{EA}$ can also be used as an approximation of the objective function defined in Equation (6) (obviously, the centroid of a cluster of trees is not necessarily a consensus tree of the cluster; furthermore, it is not necessarily a phylogenetic tree).

*Ball-Hall index adapted for tree clustering*

Another relevant criterion to consider in this review is the Ball-Hall index. In 1965, Ball and Hall (*BH*) introduced the ISODATA procedure to measure the average dispersion of groups of objects with respect to the mean square root distance, i.e. the intra-group distance. Unlike the *CH* index, the *BH* index can be used to find solutions consisting of a single consensus tree. Tahiri et al. (2022) adapted the *BH* criterion for tree clustering with *k*-means, which led to the following formula:

$$BH = \frac{1}{K} \sum_{k=1}^{K} \frac{1}{N_k} \sum_{i=1}^{N_k} RF(T_k^{maj}, T_{ki}). \tag{18}$$

Furthermore, the following formula can be used to avoid the majority-rule tree calculation:

$$BH = \frac{1}{K} \sum_{k=1}^{K} \frac{1}{N_k^2} \sum_{i=1}^{N_k-1} \sum_{j=i+1}^{N_k} RF(T_{ki}, T_{kj}). \tag{19}$$

*Silhouette index adapted for tree clustering*

The next popular criterion we consider here is the Silhouette (*SH*) width index (Rousseeuw 1987). Traditionally, the Silhouette width of cluster *k* is defined as follows:

$$s(k) = \frac{1}{N_k} \left[ \sum_{i=1}^{N_k} \frac{b(i) - a(i)}{max(a(i), b(i))} \right], \tag{20}$$

where $N_k$ is the number of elements belonging to cluster *k*, *a(i)* is the average distance between element *i* and all other elements belonging to cluster *k*, and *b(i)* is the smallest, over-all clusters *k'* different from *k*, of all average distances between *i* and all the elements of cluster *k'*.

Equations (21) and (22) can be used to calculate *a(i)* and *b(i)*, respectively, in case of tree clustering:

$$a(i) = \frac{\sum_{j=1}^{N_k} RF(T_{ki}, T_{kj})}{N_k}, \tag{21}$$

$$b(i) = \min_{1 \leq k' \leq K, k' \neq k} \frac{\sum_{j=1}^{N_{k'}} RF(T_{ki}, T_{k'j})}{N_{k'}}, \tag{22}$$

where $T_{k'j}$ is tree *j* of cluster *k'*, such that $k' \neq k$, and $N_{k'}$ is the number of trees in cluster *k'*.

The optimal number of clusters, *K*, corresponds to the maximum average value of *SH* that is calculated as follows:

$$SH = \bar{s}(K) = \sum_{k=1}^{K} \frac{[s(k)]}{K}. \tag{23}$$

The value of the *SH* index defined by Equation (23) is located in the interval between -1 and +1.

*Gap statistic adapted for tree clustering*

The last criterion that we are discussing here is the *Gap* statistic (Tibshirani et al. 2001). As the *BH* index, *Gap* allows solutions consisting of a single consensus tree. The formulas proposed by Tibshirani et al. (2001) are based on the properties of the Euclidean distance. In the context of tree clustering, Tahiri et al. (2022) adapted the *Gap* statistic by defining the total intracluster distance, $D_k$, characterizing the cohesion between the trees belonging to the same cluster *k*, as follows:

$$D_k = \sum_{i=1}^{N_k} \sum_{j=1}^{N_k} RF(T_{ki}, T_{kj}). \tag{24}$$

The sum of the average total intracluster distances, $V_K$, can be calculated using the following formula:

$$V_K = \sum_{k=1}^{K} \frac{1}{2N_k} D_k. \tag{25}$$

The *Gap* statistic, which reflects the quality of a given clustering solution with *K* clusters, is traditionally defined as follows:

$$Gap_N(K) = E_N^*\{\log(V_K)\} - \log(V_K), \tag{26}$$

where $E_N^*$ denotes expectation under a sample of size *N* from the reference distribution. The following formula (Tibshirani et al. 2001) for the expectation of $log(V_K)$ was used in our method:

$$E_N^*\{\log(V_K)\} = \log\left(\frac{Nn}{12}\right) - \left(\frac{2}{n}\right)\log(K), \tag{27}$$

where *n* is the number of tree leaves. The largest value of the *Gap* statistic corresponds to the best clustering.

## Example of application to evolutionary data

Aminoacyl-tRNA synthetases (aaRSs) are enzymes that attach the appropriate amino acid to their cognate transfer RNA. The structure-function aspect of aaRSs has long interested biologists (Woese et al. 2000, Godwin et al. 2018). It has been observed that the central role played by aaRSs in translation suggest that their evolutionary histories and that of the genetic code can be closely related (Woese et al. 2000). This information would make aaRS gene domain analysis a key component of tree-of-life inference (Bullwinkle and Ibba 2014, Unvert et al. 2017). Woese et al. examined the evolutionary profiles of each of the 20 standard aaRSs used by living cells to construct the evolutionary history of proteins organized into 5 groups (nonpolar aliphatic R group, nonpolar, aromatic R group, polar, uncharged R group, positively charged R group, and negatively charged R group). To conduct their famous aaRS analysis Woese et al. considered a total of 72 species from 3 main domains (Archaea, Eukarya and Bacteria), which can be represented by leaves of the related phylogenetic trees.

In our study, we used 36 aaRS phylogenetic trees (i.e. aaRS gene trees) originally constructed by Woese et al. These trees had different, but mutually overlapping, sets of leaves (in total 72 different species were considered). They are available on our GitHub repository along with our program at the following URL address: https://github.com/TahiriNadia/KMeansSuperTreeClustering. These 36 trees were used as input for our *KMeansSuperTreeClustering* algorithm (Tahiri et al. 2022). Our supertree clustering algorithm was carried out with the following options: the Caliński-Harabasz (Caliński and Harabasz 1974) cluster validity index was used to select the best number of clusters (the number of clusters varied from 2 to 10 in our experiments) and the penalization parameter $\alpha$ was set to 1.

In these settings, our algorithm found that the best solution for these data corresponds to a 2-cluster partitioning. Each of these clusters of trees can be represented by its own supertree. The first obtained cluster includes 19 trees for a total of 61 different species, while the second obtained cluster includes 17 trees for a total of 56 species. The supertrees (see Figures 3 and 4) for the two obtained tree clusters were inferred using the CLANN program (Creevey and McInerney 2005). In CLANN, we used the most similar supertree (dfit) method (Creevey et al. 2004) with the mrp criterion. This criterion involves a matrix representation based on the parsimony criterion. Next, we inferred the most common (by cluster) horizontal gene transfers (HGT) that characterize the evolution of phylogenetic trees included in the two obtained clusters of trees. The HGT detection method by Boc et al. (2010) was used for this purpose. It proceeds by reconciliation of the species and gene phylogenetic trees. In our case, the two obtained supertrees played the role of gene trees, while the species phylogenetic trees followed the NCBI taxonomic classification (see https://www.ncbi.nlm.nih.gov/Taxonomy/CommonTree/wwwcmt.cgi); they are presented by full edges in Figures 4 and 5. These supertrees were not fully resolved (i.e. the first supertree, see Fig. 4 contains 9 internal nodes with degree greater than 3, whereas the second supertree, see Fig. 5 contains 10 internal nodes with degree greater than 3). We used the version of the HGT algorithm available on the T-Rex website (Boc et al. 2012) and Armadillo 1.1 (Lord et al. 2012) workflow platform to identify the scenarios of HGT events that reconcile each species tree with the corresponding supertree. The root of all of these trees was placed on the edge that splits the clade of Bacteria with those of Eukarya and Archaea. Two frequent horizontal gene transfers were found for the first supertree and four for the second supertree. Our results indicate that most of aminoacyl-tRNA synthetases underwent a two-way evolution. The obtained results are in line with the results of Dohm et al. (2006) and Sharaf et al. (2019) that aminoacyl-tRNA synthetases possess two versions of most tRS, one cytosolic and one mitochondrial.

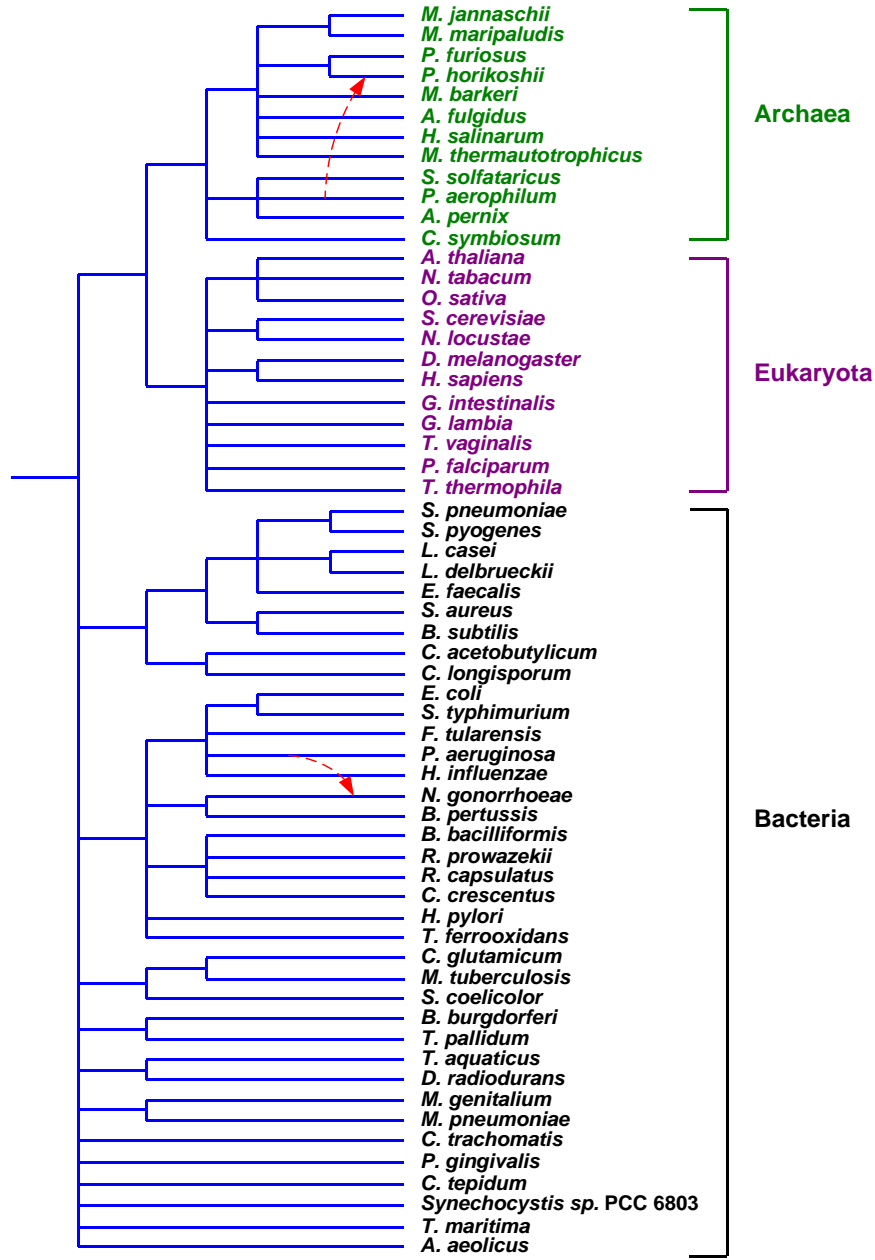

**Fig 4.** *Species tree (full edges) corresponding to the NCBI taxonomic classification constructed for 61 species from the first cluster of 19 aaRS phylogenetic trees. The two horizontal gene transfers (indicated by arrows) were found using the HGT-Detection program of Boc et al. (2012).*

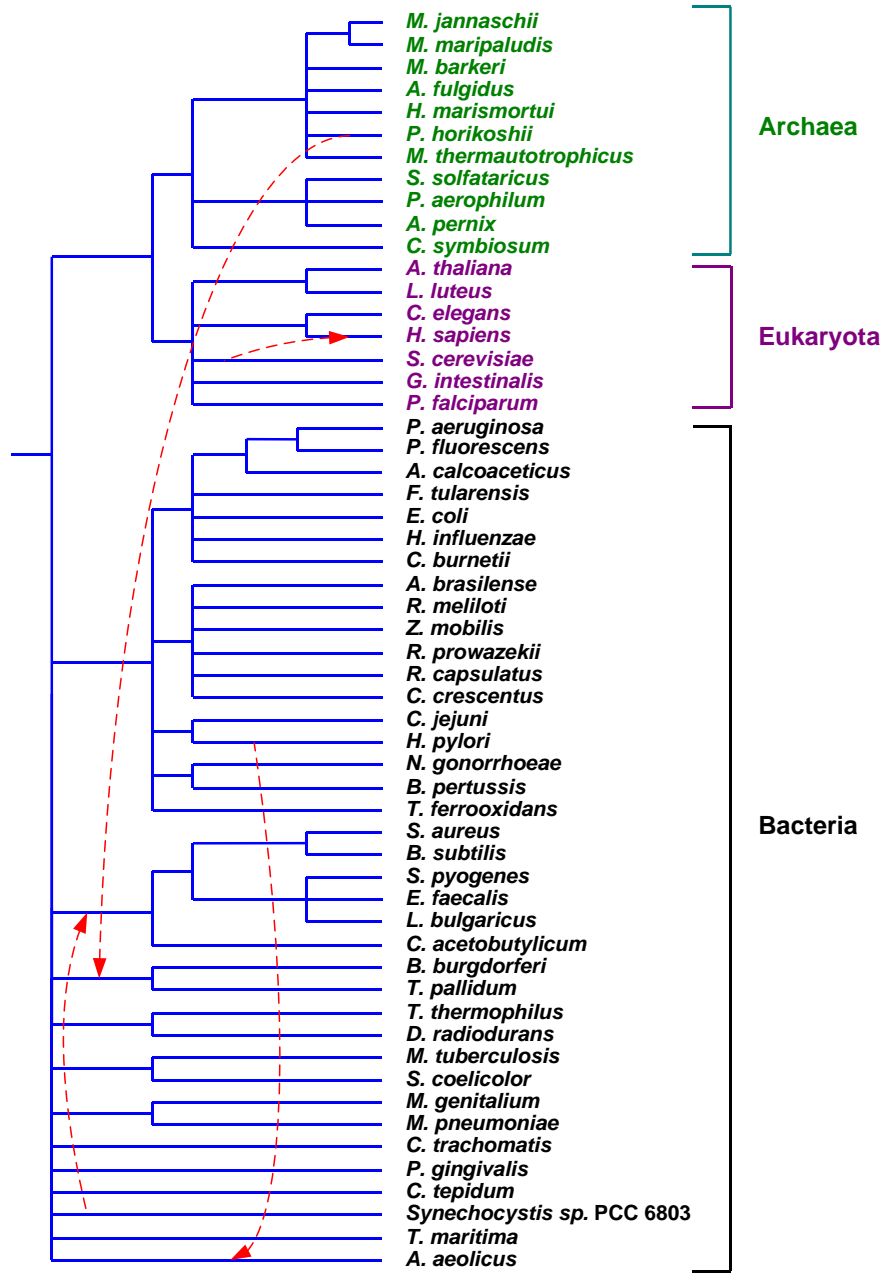

**Fig 5.** *Species tree (full edges) corresponding to the NCBI taxonomic classification constructed for 56 species from the first cluster of 17 aaRS phylogenetic trees. The four horizontal gene transfers (indicated by arrows) were found using the HGT-Detection program of Boc et al. (2012).*

## Conclusion

In this paper, we have reviewed the state-of-the-art systematic methods for inferring multiple alternative consensus trees and supertrees from a given set of phylogenetic trees (i.e. additive trees, evolutionary trees or *X*-trees). Most of the reviewed papers describe algorithms proceeding by *k*-means or *k*-medoids clustering of tree topologies. In the case of consensus tree clustering problem, all the trees should be defined on the same set of taxa (i.e. species associated to the tree leaves), whereas in the case of supertree clustering problem, the trees can be defined on different, but mutually overlapping, sets of taxa. In many instances, multiple consensus trees and supertrees represent more relevant evolutionary models than traditional single consensus trees and supertrees. The resolution of multiple consensus trees and supertrees is generally much better than that of single consensus trees or supertrees inferred by conventional methods (Maddison 1991). Thus, multiple consensus trees and supertrees have the potential of preserving much more plausible information from a set of given gene trees. Clustering seems to be an intuitive natural solution for inferring multiple consensus trees and supertrees. Tree clustering has a direct practical application in evolutionary studies. It allows one to identify sets of genes that have been affected to the same horizontal gene transfer, hybridization, intragenic/intergenic recombination events, or those that have undergone the same ancient gene duplications and gene losses during their evolution (Makarenkov et al. 2004, Bapteste et al. 2004, Diallo et al. 2006, Boc and Makarenkov 2011).

Since the beginning of the Tree of Life inference project (Maddison et al. 2007), the number of studies dealing with supertree theory has grown considerably. The methods described in this paper can be used for inferring multiple alternative subtrees of the Tree of Life as it contains many unresolved clades (i.e. subtrees with high degrees of its internal nodes). From the practical point of view the problem of constructing multiple alternative supertrees is more relevant than that of constructing multiple alternative consensus trees because most of currently available gene trees are not defined on exactly the same sets of taxa. However, to the best of our knowledge, the only study addressing this relevant problem remains the recent work of Tahiri et al. (2022). The authors of this work showed how some remarkable properties of the Robinson and Foulds topological distance (original or normalized) and the *k*-means partitioning algorithm can be used to achieve very promising tree clustering performance. Finally, in the application section, we showed how this method can be applied to cluster phylogenetic trees from the famous aaRS phylogenetic dataset originally described by Woese et al. (2000).

An interesting option for further investigations consists in the use of some other popular tree distances in the objective function of clustering algorithms. Among them, we need to mention the branch score distance (Kuhner and Felsenstein 1994) and the quartet distance (Bryant et al. 2000), which also have the Euclidean properties as the square root of the Robinson and Foulds distance.